____________________

# Debonding of a soft adhesive fibril in contact with an elastomeric pillar

Aymeric Duigou-Majumdar,[ab] Pierre-Philippe Cortet[*a] and Christophe Poulard [*b]

The debonding criterion of fibrils of soft adhesive materials is a key element regarding the quantitative modelisation of pressure sensitive adhesive tapes' peeling energy. We present in this article an experimental study of the detachment of a commercial acrylic adhesive tape from the top surface of a single micrometric pillar of PDMS elastomer. During an experiment, the pillar and the adhesive, after being put in contact, are separated at a constant displacement rate, resulting in the formation, the elongation and the final detachment of a fibril of adhesive material. A systematic study allows us to uncover power laws for the maximum force and the critical elongation of the fibril at debonding as a function of the diameter of the cylindrical pillar which controls the diameter of the fibril. The scaling law evidenced for the critical elongation appears as a first step toward the understanding of the debonding criterion of fibrils of soft adhesive materials. In addition, viscoelastic digitation at the triple debonding line is observed during detachment for large pillar diameters. The wavelength and penetration length of the fingers that we report appear to be consistent with existing models based on pure elastic mechanical response.

## 1 Introduction

Building a physical model able to predict the adherence surface energy $\Gamma$ (*i.e.*, the energy release rate) of pressure sensitive adhesive tapes (PSA) during their peeling from a substrate has focused a lot of attention since the 1950's.[1] During debonding of a PSA, the thin layer of nearly incompressible soft adhesive material confined between the rigid substrate and the backing tape (in peeling experiments) or the rigid flat punch (in probe-tack experiments) generally first experiences the nucleation and the growth of cavities[2,3] or a fingering instability of the debonding front.[3,4] In both cases, these processes lead to a second stage of the detachment where fibrils of adhesive material are stretched until debonding from the substrate.[2–10] Another important guide in the modelling of the peeling surface energy of PSA is the fact that its dependence with the peeling rate and the temperature has been shown to follow the same time-temperature equivalence as the small-strain visco-elasticity of the adhesive material of the PSA.[6,11–14]

An early proposed theoretical strategy[11,12,15] to describe the peeling energy of PSA consists in saying that it is driven by the work to stretch the fibrils of adhesive material up to debonding. In such an approach, the peeling surface energy can more precisely be written as $\Gamma = e\int_0^{\varepsilon_c} \sigma(\varepsilon, \dot\varepsilon) d\varepsilon$, where $e$ is the initial thickness of the adhesive layer, $\sigma(\varepsilon, \dot\varepsilon)$ the strain-rate dependant stress–strain curve of the adhesive material in uniaxial stretching, $\varepsilon$ the strain of the adhesive material and $\varepsilon_c$ its critical value at debonding, *i.e.* at the peeling front. In this expression, when the peeling front advances, the local strain rate $\dot\varepsilon$ is parameterised by the strain $\varepsilon$ going from 0 to $\varepsilon_c$, from the time the adhesive material enters the process zone ahead of the peeling front to the moment it debonds. In this framework, the interfacial energy between the adhesive layer and the substrate enters the problem only indirectly in the determination of the critical strain at debonding of the fibrils of adhesive material $\varepsilon_c$.

Several works have recently given experimental credit to this viewpoint. Villey *et al.*[8] showed that two acrylic PSA having the same linear visco-elasticity but different large-strain rheology present different peeling energies in relation with their difference in the maximum extensibility of the fibrils of adhesive material at debonding. This work evidenced the crucial role of large deformations on the peeling energy which cannot be accounted for by the small-strain visco-elasticity of the adhesive. More recently, Chopin *et al.*[16] succeeded to render such model semi-predictive for the same acrylic PSA: they characterise the large deformation stress–strain elongation rheology $\sigma(\varepsilon, \dot\varepsilon)$ of the adhesive material as a function of the strain rate $\dot\varepsilon$. Coupling these rheological measurements to the actual values of the critical elongation $\delta_c = e\varepsilon_c$ of the fibrils of adhesive material at debonding experimentally determined during the peeling experiments, they succeeded to predict values of the energy release rate $\Gamma$ in good agreement with the direct measurements (by peeling force measurements). Chopin *et al.*[16] also explained how the time-temperature equivalence of the small-strain rheology of the adhesive material is recovered in the peeling energy $\Gamma$: this is a consequence, at least for the PSA considered in this study, of the fact that the large-strain rheology dependences with the strain rate and the temperature can be encoded in a unique multiplicative prefactor matching the small-strain elastic modulus.

[a] *Université Paris-Saclay, CNRS, FAST, 91405, Orsay, France. E-mail: pierre-philippe.cortet@universite-paris-saclay.fr*
[b] *Université Paris-Saclay, CNRS, Laboratoire de Physique des Solides, 91405, Orsay, France. E-mail: christophe.poulard@universite-paris-saclay.fr*

One understands here that the missing element to establish a predictive model for the peeling energy of PSA is the criterion determining at which elongation a fibril of adhesive material debonds. Several criteria have been introduced in the literature in order to predict the peeling energy $\Gamma$ of different types of PSA: for example, Kaelble[11,15] and Gent and Petrich[12] introduced a critical stress $\sigma_c$ and Derail *et al.*[14] a constant critical stretch. In order to model the enhancement of the peel energy due to a texturation of the elastic substrate by an array of microscopic pillars, Poulard *et al.*[17] used a critical force for the detachment of the adhesive fibrils from the top of the pillars which is proportional to the pillar diameter, a criterion inspired by the JKR theory of sphere-plane contacts.[18] However, as highlighted by Chopin *et al.*,[16] no real understanding of the debonding criterion of a stretched microscopic fibril of adhesive material is currently available.

This is the question we tackle in this article, from an experimental point of view and for the case of acrylic adhesives used in standard office tapes. Putting into contact a layer of an acrylic PSA deposited on a glass lens with a micrometric cylindrical pillar of PDMS before separating them, we trigger the formation of a unique fibril of adhesive material of controlled diameter. We are then able to study the evolution of the force and of the elongation of the fibril until debonding. By varying the initial fibril diameter, we observe power law relationships for both the maximum force and the maximum elongation at debonding. The typical wavelength and penetration length of a digitation instability appearing at the triple line is also analysed. These are in good agreement with existing models dealing with digitation instability for purely elastic contacts. Finally, the consistency of the observed power laws with the mechanical response of the system is used to propose a debonding criterion for a fibril of adhesive material. This criterion consists in a critical elongation of the adhesive material fibril proportional to the square root of the diameter of the fibril times a characteristic microscopic length whose physical meaning remains to be understood.

## 2 Sample preparation and experimental setup

### 2.1 Sample preparation and properties

We study the contact between a commercial acrylic adhesive tape and a cylindrical pillar of a silicone elastomer (PDMS) on the top of a flat layer of the same material. The substrate of PDMS is produced as follows. A cylindrical cavity of diameter $d$ (ranging from 30 to 5000 µm) is carved in a 28 µm-thick layer of an epoxy photoresist resin (SU8-2025 from Kayaku Advanced Materials) using an optical lithography machine (MicroWriter ML 3® from Durham Magneto Optics Ltd). The PDMS susbtrate is obtained by pouring a solution of Sylgard™ 184 from Dow chemical® (using a solution with a 10:1 w/w ratio of elastomer to cross-linker solution) in the epoxy resin mould which is then left to crosslink at room temperature during 3 days before it is peeled from the mould. The layer of PDMS elastomer (typically 1 mm thick) with the pillar on the top is finally fixed on a silicium wafer. The PDMS substrate has a $E_s = 1.3 \pm 0.1$ MPa elastic modulus and could be considered as a purely elastic linear material.[19] The surface free energy of the crosslinked Sylgard 184 is $\gamma_s = 21.5 \pm 0.1$ mN m$^{-1}$.[20]

The adhesive used is the commercial adhesive tape 3M Scotch® 666. This double sided tape is composed of two layers of acrylic adhesive of thickness $22 \pm 2$ µm separated by a $37 \pm 1$ µm thick film of UPVC. The adhesive layers on the faceside and the backside of the tape are not made of the same material. The layer of interest in this study, *i.e.* the one in contact with the PDMS pillar, is always the backside.

We measured the surface tension of this backside adhesive layer following the "Owens, Wendt, Rabel and Kaelble" method[21] consisting in depositing sessile droplets of different liquids on the material surface. The obtained surface free energy of the adhesive layer is $\gamma_a = 25 \pm 1$ mN m$^{-1}$. In addition, by depositing a droplet of uncrosslinked Sylgard 184 on the 3M666 backside surface, we measured a static contact angle of $\theta = 30 \pm 3°$. From this contact angle and the values of measured surface tensions $\gamma_s$ and $\gamma_a$, we can compute the interfacial energy between the adhesive layer and the Sylgard 184 PDMS, $\gamma_{as} = \gamma_a - \gamma_s \cos\theta = 6 \pm 1$ mN m$^{-1}$, which value is in order of magnitude classical for a polymer/polymer surface free energy. We can also obtain the work of adhesion $W_0 = \gamma_s + \gamma_a - \gamma_{as} = \gamma_s(1 + \cos\theta) = 40 \pm 1$ mN m$^{-1}$ of the interface between Sylgard 184 and the 3M666 tape backside surface.

In order to characterise the linear visco-elasticity of the adhesive tape, we realised shear rheometer measurements with an Anton Paar MCR 301 device in the parallel plate geometry (plate diameter 5 cm) on a stacking of 10 layers of 3M Scotch® 666 tape. In Fig. 1, we report the evolution of the storage $\mu'$ and loss $\mu''$ shear modulus measured at a temperature of 22 °C, for a shear strain amplitude $\gamma$ of 1% of the material thickness and in the angular frequency range $4 \times 10^{-2}$ rad s$^{-1}$ < $\omega$ < $10^2$ rad s$^{-1}$. The measured modulus are representative of the mean effective behavior produced by the two sides of the tape. They have nevertheless been corrected to account for the fact the 37 µm thick carrier tape is rigid. The order of magnitude and dependencies reported in Fig. 1 are in good agreement with other measurements done on similar acrylic adhesive tapes.[8,16] In the experiments presented in the following, the strain rate at which the tape backside adhesive layer will be stretched is typically $\dot{\varepsilon} \sim 0.05$ s$^{-1}$. The corresponding angular frequency, $\omega = \dot{\varepsilon}/\gamma \sim 5$ rad s$^{-1}$, is associated in Fig. 1 to the following values of the storage modulus $\mu' = 6.6 \times 10^4$ Pa and loss modulus $\mu'' = 2.6 \times 10^4$ Pa. The corresponding elastic modulus is $E = 2\mu'(1 + \nu) \simeq 3\mu' \simeq 2.0 \times 10^5$ Pa since the adhesive material is nearly incompressible ($\nu \simeq 0.5$).[1]

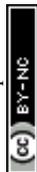

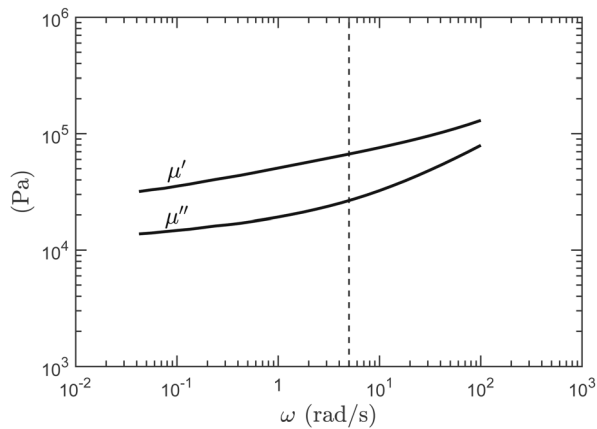

Fig. 1 Storage $\mu'$ and loss $\mu''$ shear modulus of the commercial adhesive tape 3M Scotch® 666 as a function of the angular frequency $\omega$ measured at a temperature of 22 °C for a shear strain amplitude of 1% of the material thickness.

### 2.2 Experimental setup

Experiments consist in contacting vertically a plano-convex glass lens, covered with a layer of 3M Scotch® 666 tape, with a cylindrical PDMS pillar emerging from a flat PDMS substrate. The experimental setup is sketched in Fig. 2. The lens (Edmund optics 89414), 1.5 mm high, 5 mm in diameter and with the spherical convex face having a radius of curvature of $R = 11.5$ mm, is fixed on a 25 mm square glass plate. Before each experiment, a fresh piece of adhesive tape is placed on the spherical face of the lens. The adhesive layer, typically of 2 cm size, completely covers the glass lens and part of the glass plate on which the lens is fixed. The interfacial energy between the glass lens and the face-side of the adhesive tape is much larger than the one between the PDMS and the back-side of the adhesive tape. In addition, the contact of the adhesive tape with the glass lens and the glass plate is always much larger than the one with the PDMS pillar. This ensures that, during debonding, the breakage will occur only at the interface between the adhesive tape and the PDMS pillar. The pillar is carefully centered with respect to the lens using a XY-axis horizontal stage. The lens is rigidly coupled to a vertical linear stage (PI M-112.1DG1) allowing us to control its relative vertical position $\Delta z$ with a nanometric resolution. The height $h$ of the cylindrical pillar is of 28 µm and its diameter $d$ is ranging from 30 µm to 5000 µm. On the other hand, the PDMS substrate is mounted on a flexible double-plate of stiffness $K = 1390 \pm 10$ N m$^{-1}$ coupled to a capacitive position sensor (PI D-100.00, nanometric resolution) providing together a measure of the contact force $F$ with a resolution of the order of the µN. This force measurement allows us to compute the actual displacement $\delta$ ($=\Delta z - F/K$) of the lens relative to the substrate, by subtracting the deformation of the force gauge. The noise on the force measurement is of ±0.15 mN. Parallelism and alignment are ensured by the use of two goniometers placed perpendicular to each other. A top view camera, coupled with a telescopic objective, allows us to visualise the contact between the adhesive tape and the top of the pillar during the approach and detachment processes (see a typical image in inset of Fig. 2).

The curvature of the glass lens ensures that no other contact than the one on the pillar is created. For low values of the pillar diameter $d$, the curvature of the lens will be negligible and the pillar will act like a flat punch on the adhesive material when the contact proceeds. For example, for the lowest pillar diameter $d = 30$ µm, the difference in separation distance between the adhesive layer and the center or the periphery of the pillar top surface is of $\sim 0.01$ µm (in theory only, because of surfaces roughness). For $d = 500$ µm, the difference in the separation distance at the center and at the periphery of the pillar is of $\sim 3$ µm. For larger pillar diameters $d$, the lens curvature will rapidly not be negligible. The contact area will finally be smaller than the pillar top surface. It will be a sphere-plane contact sandwiching an adhesive layer of thickness ($e = 22$ µm) much smaller than the lens radius of curvature. This contact might in principle be described by the JKR theory of adhesive sphere-plane contacts.[18] We will however see that, in this regime, the debonding of the adhesive layer from the PDMS proceeds through a fingering instability of the debonding front,[4,19,22–24] revealing a richer process.

Experiments are performed at a temperature of 22 ± 2 °C. The progress of a typical experiment consists first in the downward motion of the lens covered with the adhesive tape, approaching the PDMS pillar at velocity 0.5 µm s$^{-1}$. The total downward distance travelled by the stage is set to typically overshoot the contact with the pillar of a few micrometers. From the visualisation by the camera, we identify the moment at which contact is realised and can therefore precisely measure the distance $I$ travelled by the stage during the final "indentation" phase of the downward motion. This phase can be observed on the left panel of Fig. 3 showing the contact force $F$ as a function of the position $\Delta z$ of the translation stage for an experiment with a pillar of diameter $d$ = 550 μm.

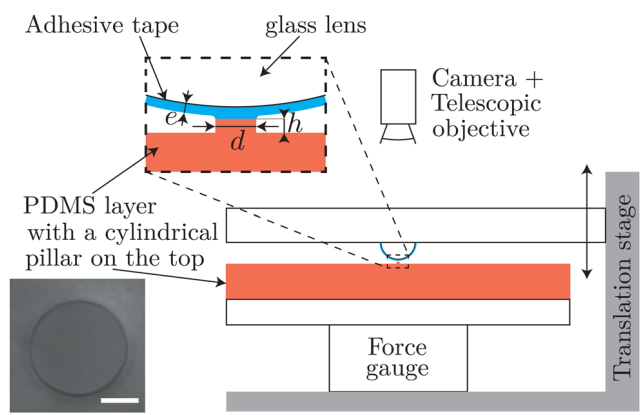

Fig. 2 Scheme of the experimental setup. An adhesive material layer of thickness $e = 22$ µm is placed on a plano-convex glass lens of radius $R = 11.5$ mm. The lens is coupled to a vertical translation stage. In front of the lens is placed a PDMS horizontal layer from which is emerging a vertical PDMS cylindrical pillar of diameter $d \in [30:5000]$ µm and height $h = 28$ µm. The pillar and lens axes of revolution are aligned precisely. The PDMS substrate is supported by a double-plate flexible sensor measuring the vertical force when the lens covered by the adhesive enters into contact with the PDMS pillar. The contact is visualised from the top using a camera coupled to a telescopic objective. A typical image is shown at the bottom left of the figure in the case of a pillar of diameter $d = 250$ µm (the white line has a 100 µm length). This image has been taken at the very beginning of the separation phase of the experiment, a moment at which the adhesive layer and the top of the PDMS pillar are in complete contact.

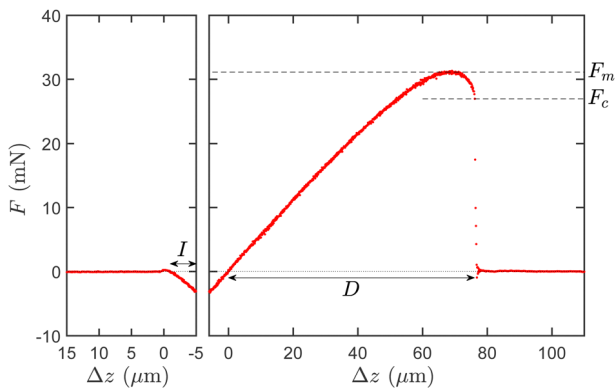

Fig. 3 Contact force $F$ as a function of the vertical position $\Delta z$ of the translation stage holding the glass lens covered with the adhesive tape for an experiment with a PDMS pillar of diameter $d$ = 550 μm. Left panel: Approach and contact at a velocity of 0.5 μm s$^{-1}$. Right panel: Separation phase during which the lens is going up at velocity $V$ = 1 μm s$^{-1}$. In the left panel, we define the indentation distance $I$. In the right panel, we define the maximum detachment force $F_m$, the critical detachment force $F_c$ and the displacement at debonding $D$.

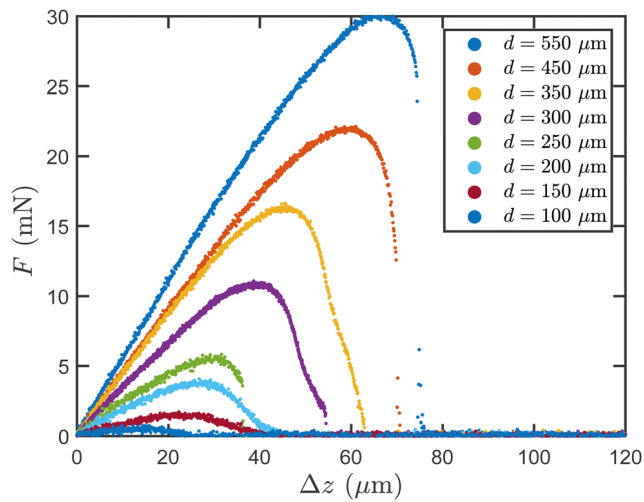

Fig. 4 Examples of force–displacement curves for pillar diameters $d$ between 100 μm and 550 μm during the separation phase.

In the example of Fig. 3, $I$ = 4.82 μm and the corresponding force $F_I$ = −3.08 mN such that the effective indentation of the adhesive layer by the PDMS pillar is $\delta_I = I + F_I/K$ = 2.60 μm. With our protocol, since the initial distance between the adhesive tape and the PDMS substrate before the contact is not known precisely, we do not control the value of the indentation $\delta_I$ of the adhesive layer by the PDMS pillar. We nevertheless measure it precisely. In the experiments we report in the following, the indentation $\delta_I$ will typically vary between vanishing values and 20 μm. In order to restrict our analysis to contacts of the adhesive with the pillar only, we excluded experiments with larger values of $\delta_I$: for indentations approaching the height $h$ = 28 μm of the PDMS pillar, it is indeed frequently observed that the adhesive material enters in contact with the flat base of PDMS from which the pillar emerges.

We finally break the contact by moving upward the lens covered with the adhesive at controlled velocity $V$ = 1 μm s$^{-1}$. The typical strain rate at which it is stretched is $\dot{\varepsilon} \sim V/e \simeq 0.05$ s$^{-1}$. The right panel of Fig. 3 reports the evolution of the contact force $F$ as a function of $\Delta z$ during this separation phase, for the same experiment as for the contact phase of the left panel. From the force–displacement curves, we define the maximum force $F_m$ during the separation. The displacement $D$ to reach the complete debonding, identified from the visualisation of the contact, is also reported and corresponds well to the moment at which the contact force $F$ drops to zero. Note that the stage reference position $\Delta z$ = 0 used to compute the debonding displacement $D$ has been set to the position at which the force crosses zero during the separation phase of the experiment.

In Fig. 4, we report several examples of force–displacement curves for pillar diameters $d$ between 100 μm and 550 μm. For the experiments with pillar diameters $d$ larger than 450 μm (as it is the case in Fig. 3), we observe that the final phase of the detachment proceeds via an instantaneous drop of the force from a critical detachment force $F_c$ to zero. For experiments at pillar diameters $d$ lower than 100 μm, no brutal drop of the force can be detected during the detachment and $F_c$ is set to zero. For $d$ in between, both behaviors can be observed. For two diameters (200 and 350 μm), no force drop is observed for the set of experiments we have conducted. In the following, for this intermediate range of pillar diameter, we compute the average value of the critical force $F_c$ for each pillar diameter $d$ considering only the subset of experiments for which the critical force is non-zero.

Finally, we introduce the observable $\delta = \Delta z - F/K$ which measures the elongation of the system "substrate + adhesive". In particular, we define the critical elongation as $\delta_c = D - F_c/K$ which is a measure of the substrate-adhesive joint elongation "just before" debonding.

## 3 Experimental results

### 3.1 Maximum force and critical elongation

In Fig. 5, we report, in log–log scale, the maximum force $F_m$ during the detachment process as a function of the indentation $\delta_I$ for all experiments with pillar diameters $d \geq 100$ μm. One first observes that there is a statistical distribution of the values of the maximum force when several experiments realised with close control parameters $\delta_I$ and $d$ are compared. This is the consequence of the fact that a contact between a PDMS pillar and the layer of adhesive tape deposited on the glass lens is never strictly identical to the other. Further in our study, we will consequently consider the ensemble average of the maximum force (and of other observables of interest) over several experiments in order to reveal the mean behavior of the system.

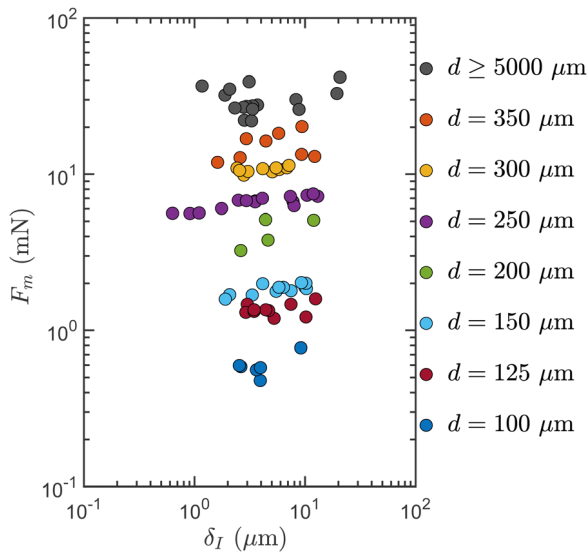

Fig. 5 Maximum force $F_m$ during the detachment process as a function of the indentation $\delta_I$ for all experiments with $d \geq 100$ μm, in log–log scale. Each marker color corresponds to a pillar diameter $d$. Data points corresponding to $d \geq 550$ μm gather all the experiments for pillar diameters $d$ = 550, 1000, 2000 and 5000 μm.

Beside, for $d \leq 70$ μm, it has proven impossible to extract a robust value of the maximum force for each experiment since the force signal is of the same order or even lower than the noise on the force measurement which is typically of ±0.15 mN.

In Fig. 5, for 100 μm $\leq d \leq$ 450 μm, we observe that $F_m$ is a very weakly increasing function of $\delta_I$. In parallel, we see that the maximum force $F_m$ increases regularly with the pillar diameter $d$ before a saturation is observed beyond $d$ = 450 μm. For pillar diameters $d$ larger than 550 μm, $F_m$ is indeed both independent of the indentation $\delta_I$ and of the pillar diameter $d$.

These observations are confirmed when plotting in Fig. 6, the average maximum force $\langle F_m \rangle$ as a function of the pillar diameter $d$. Here, we are able to report data for pillar diameters down to $d$ = 30 μm by computing the ensemble average of the force–separation curves $F(\delta)$ over all the experiments at a given pillar diameter (see Fig. 7). For $d$ = 30 μm in particular, we realised 131 experiments, the ensemble average of which leads to a reasonably smooth force–separation curve cleansed of most of the measurement noise. We finally extract the maximum force $\langle F_m \rangle$ from the ensemble-averaged force–separation signal just as we previously made for individual experiments. For the experiments with $d \geq 100$ μm, for which we can compute both the maximum force $\langle F_m \rangle$ of the ensemble-averaged force–separation signal and the ensemble average of the maximum force of individual force signals, we verify that the relative difference between the two observables is always less than 5%. As an illustration, we show in Fig. 7 the superposition of the force–separation curves for the 12 experiments conducted at $d$ = 70 μm and of their ensemble average. One can observe here the strong reduction of the experimental noise resulting from the ensemble average which finally leads to a $F(\delta)$ curve smooth enough to be analysed.

In Fig. 6, for $d \leq 450$ μm, the maximum force $\langle F_m \rangle$ is increasing with the pillar diameter $d$ following a power law of exponent 5/2 over more than a decade of pillar diameter. Increasing the pillar diameter above 450 μm, the maximum force is slowly tending toward its asymptotic value $F_\infty$ = 38.2 ± 8 mN measured for a flat PDMS substrate (corresponding to $d \to \infty$). The pillar diameter at the cross-over between the two regimes is $d_c$ = 500 ± 40 μm. Fig. 6 also shows the average critical force at debonding $\langle F_c \rangle$. As already mentioned, $\langle F_c \rangle$ is set to zero, for the diameters $d$ for which a brutal drop of the force has never been detected at the end of the detachment process. For diameters $d$ larger than 100 μm (and except for $d$ = 200 and 350 μm), the average critical force $\langle F_c \rangle$ is slightly lower than and, to the first order, is proportional to the maximum force with $\langle F_c \rangle / \langle F_m \rangle = 0.75 \pm 0.15$.

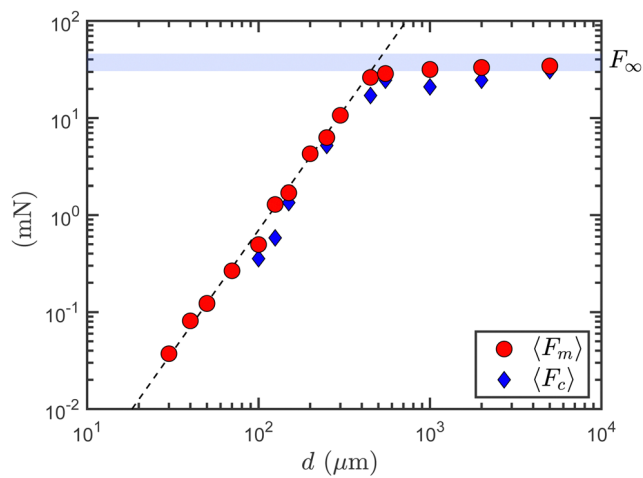

Fig. 6 Average maximum force $\langle F_m \rangle$ and critical force $\langle F_c \rangle$ as a function of the pillar diameter $d$. The blue horizontal band shows the value $F_\infty$ = 38.2 ± 8 mN of the average maximum force measured for a flat PDMS substrate, corresponding to $d \to \infty$ (the band thickness shows the dispersion on $F_\infty$). The dashed line shows a power law of exponent 5/2.

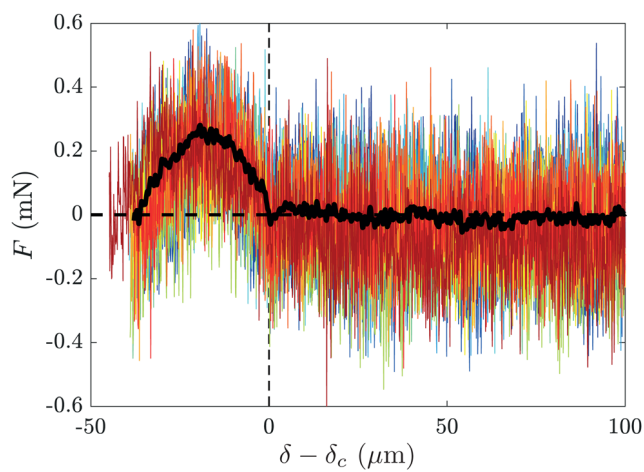

Fig. 7 Contact force $F$ as a function of the system elongation $\delta - \delta_c$ for the 12 experiments realised with a pillar of diameter $d$ = 70 μm (colored curves). The black thick line corresponds to the force–separation curve obtained by ensemble average over the 12 experiments realised at $d$ = 70 μm.

In Fig. 8, we report, as a function of the pillar diameter $d$, the average critical elongation at debonding $\langle \delta_c \rangle$, which is extracted from the force–separation curves ensemble-averaged over all experiments at a given pillar diameter $d$. The ensemble averages of $F(\delta)$ again allow us to report data down to the smallest considered pillar diameter, $d = 30$ μm. In Fig. 8, even if the data points are much more scattered, the behavior of $\langle \delta_c \rangle$ is qualitatively similar to the one observed for the maximum force $\langle F_m \rangle$: a power-law behavior for $d \leq d_c \simeq 500$ μm and a plateau for $d \geq d_c$. An important difference however is the fact that the increase of $\langle \delta_c \rangle$ with $d$ at small pillar diameter is much slower than for the maximum force: it is indeed well accounted for by the power law $\langle \delta_c \rangle = (\ell_0 d)^{1/2}$ with a characteristic length $\ell_0 = 10 \pm 2$ μm. At large pillar diameters ($d \geq d_c \simeq 500$ μm), the critical elongation $\langle \delta_c \rangle$ saturates to a constant value matching the critical elongation at debonding measured for a flat PDMS substrate, $\delta_\infty = 70.3 \pm 9.3$ μm.

It is worth to remember that the critical elongation $\delta_c$ includes both the deformation of the adhesive material and the deformation of the PDMS pillar. We can however estimate the elongation of the PDMS pillar at debonding: it is indeed directly related to the critical force $F_c$ via the elastic relation $\delta_{\text{pillar}} = 4 F_c h / (\pi d^2 E_s)$, where $h = 28$ μm is the pillar height at rest. We have seen that at low pillar diameter, below 100 μm, the critical force $F_c$ is zero and that at larger pillar diameter it is equal in ensemble average to $(0.75 \pm 0.10) \langle F_m \rangle$ where $\langle F_m \rangle$ is the ensemble-averaged maximum force during debonding. Recalling that, for pillar diameters below $d_c \simeq 500$ μm, $\langle F_m \rangle \simeq F_\infty (d/d_c)^{5/2}$ with $F_\infty \simeq 38.2$ mN, we get the following upper bound $\delta_{\text{pillar}}/\delta_c = 0.75 \times 4 F_\infty h / (\pi d_c^{5/2} \ell_0^{1/2} E_s) \simeq 4\%$ for the ratio of the pillar elongation to the total elongation of the system at debonding. This estimate allows us to state that nearly all the elongation $\delta_c$ is realised in the adhesive material. Beside, in Fig. 8, one sees that starting from the lowest pillar diameter $d = 30$ μm and going to the largest, the relative maximum elongation $\delta_c/e$ of the adhesive material ranges from ~100% to ~300% of the initial thickness of the adhesive layer $e = 22$ μm.

### 3.2 Scenario of the debonding process

Before discussing the physical meaning of the previous results, it is important to describe more precisely the scenario of the debonding process depending on the pillar diameter $d$. To illustrate this discussion, we report in Fig. 9 a series of images of the contact recorded by the camera for five experiments at pillar diameters $d = 100, 200, 300, 450$ and 5000 μm. These images are accompanied by the corresponding stress–strain curves $\sigma$ vs. $\varepsilon$, where $\sigma = 4F/\pi d^2$ is the contact force normalised by the pillar top surface and $\varepsilon = \delta/e$ is the system elongation normalised by the initial thickness of the layer of adhesive material.

We first highlight the fact that the initial contact between the adhesive and the pillar involves the whole top surface of the pillar for all the experiments with pillar diameters $d$ lower than $d_2 = 775 \pm 225$ μm. For these experiments, the initial diameter

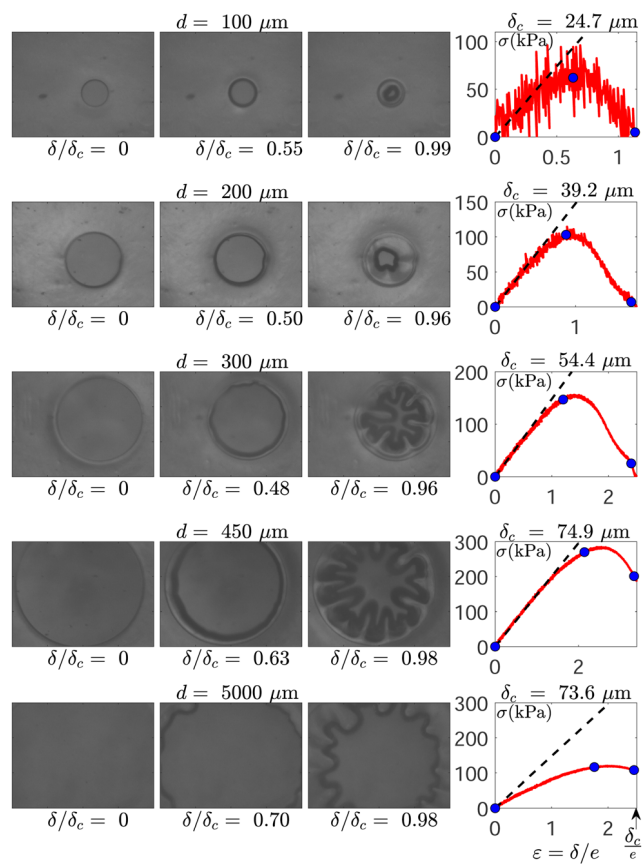

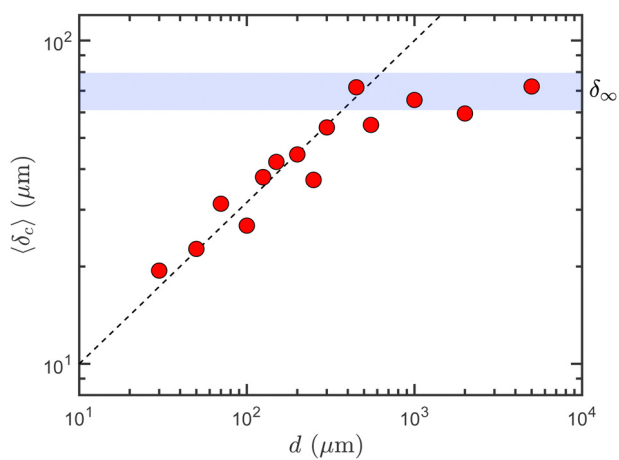

Fig. 8 Average critical elongation at debonding $\langle \delta_c \rangle$ as a function of the pillar diameter $d$. The dashed line shows a power law of exponent 1/2. The blue horizontal band shows the value $\delta_\infty = 70.3 \pm 9.3$ μm of the critical elongation at debonding measured for a flat PDMS substrate, corresponding to $d \to \infty$ (the band thickness shows the dispersion on $\delta_\infty$).

Fig. 9 Series of images of the contact during the separation phase for five experiments at different diameters: $d = 100, 200, 300, 450$ and 5000 μm from top to bottom. From left to right, images are shown for an increasing system elongation $\delta$ whose value is given normalised by the critical elongation at debonding $\delta_c$. In each row, the right panel shows the stress–strain curve $\sigma(\varepsilon)$ during separation, where the moments at which the images are picked are highlighted by blue disks. The dashed straight line has a slope of $\sigma/\varepsilon = 1.48 \times 10^5$ Pa.

$d_i$ of the circular contact is equal to $d$ and the separation phase consists in the formation, the elongation and the final debonding of a fibril of adhesive material extracted from the adhesive layer. On the contrary, for pillar diameters $d \geq d_2$, the contact diameter is always smaller than the pillar diameter because of the lens curvature ($R = 1.15$ cm) and the situation is close to a sphere-plane contact sandwiching a thin layer of adhesive material. For these experiments at $d \geq d_2$, we measure, at the beginning of the separation phase, contacts with an initial diameter $d_i$ in the range $600 \pm 100$ μm.

For all experiments at $d \leq d_2$, the diameter of the circular contact between the adhesive layer and the PDMS pillar remains equal to its initial value $d_i = d$ during a first stage of the separation process. During this stage, which typically ends when the system elongation $\delta$ reaches values of the order of 1/3rd of the critical elongation at debonding $\delta_c$, the contact stress $\sigma$ increases nearly linearly with the deformation $\varepsilon$ as can be seen in Fig. 9. Measuring the slope of the stress–strain curves $\sigma(\varepsilon)$ in this regime, we can evaluate an effective elastic modulus of the adhesive layer of $E_e = \sigma/\varepsilon = (1.48 \pm 0.20) \times 10^5$ Pa. This modulus appears to be independent of the pillar diameter $d$ and is consistent in order of magnitude with the value ($2.0 \times 10^5$ Pa) of the mean elastic modulus of the 3M Scotch® 666 tape measured with a rheometer (see Fig. 1) and which accounts for the effective behavior resulting from the joint action of the faceside and of the backside of the tape. Following this linear stage (and still focusing only on the experiments at $d \leq d_2$), the contact area begins to very slowly decrease while keeping its circular shape during the stage where the increase of the force slows down to zero and which ends when the force $F$ (or equivalently the normalised force $\sigma$) reaches its maximum value $F_m$ ($\sigma_m$, respectively). At this moment where $F = F_m$, we experimentally observe that the contact diameter has decreased by about 10% compared to the initial contact (see Fig. 9), which corresponds to a 20% decrease in contact area. Besides, the force $F_m$ is typically 20% lower than what would be expected from the extrapolation of the linear elastic behavior observed at smaller strain.

The following and final stage of the experiments, during which the force decreases, depends significantly on the value of the pillar diameter $d$. For pillar diameters lower than $d_0 \simeq 100$ μm, we observe a progressive decrease of the contact force $F$ down to zero in correlation with the progressive decrease of the contact area which keeps an approximately circular shape until complete debonding at $\delta = \delta_c$. For the experiments at pillar diameters $d$ larger than $d_1 = 225 \pm 25$ μm, the final stage is very different: a fingering instability of the initially circular debonding front begins at the moment the separation force $F$ starts to decrease. The fingers of the debonding front are then growing until they reach a maximum length $L_f$ of typically $\sim 120$ μm. At this moment, the complete debonding proceeds suddenly and is most often associated with an instantaneous drop to zero of the contact force from a finite value in the range $F_c = (0.75 \pm 0.15)F_m$, as illustrated in Fig. 9 in term of stress for the experiment at $d = 450$ μm and in Fig. 3 for an experiment at $d = 550$ μm. Nevertheless, in this range of pillar diameter $d_1 \leq d \leq d_2$, a few experiments still proceeds via a progressive decrease of the force down to zero (see the experiment at $d = 300$ μm in Fig. 9). For pillar diameters in the intermediate range $d_0 \leq d \leq d_1$, both behaviors are also observed for the force decrease depending on the experiment. The debonding front shows small perturbations with wavelengths compatible with those of the fingering instability observed for $d_1 \leq d \leq d_2$, but the small size of the contact does not allow here a significant growth of the fingers before the complete detachment.

For pillar diameters $d \geq d_2$, during the first stage of the experiments, the scenario is different and the debonding front start to progress slowly as soon as the contact force becomes positive. Obviously, this implies that an elastic modulus cannot here be measured from the slope of the stress-strain curve. The debonding front progressively accelerates while the contact area keeps a circular shape until the maximum force $F_m$ is reached. At this moment, the fingering instability starts to develop and the final scenario is similar to the one for $d_1 \leq d \leq d_2$. The growth of the fingers proceeds until they reach a length $L_f$ of the order of 120 μm. The end of the finger growth is associated to the final critical detachment and the drop of the contact force from the critical value $F_c$. We highlight that here the final distance between the tips of the opposite fingers is still of about 200 μm just before the final debonding.

The difference in the scenario of the debonding process for $d$ lower or larger than $d_2$ during the first stage of the experiment is surely related to the fact the initial adhesive-pillar contact is complete for $d \leq d_2$ and incomplete for $d \geq d_2$: the debonding front is pinned on the circular edge of the cylindrical pillar top at the beginning and during the first part of the debonding process for $d \leq d_2$ whereas it is allowed to freely advance from the beginning of the separation phase for $d \geq d_2$.

In order to characterise the fingering instability of the debonding front, we report in Fig. 10 the diameter of the contact $d_f$ at the moment the first undulations of the front emerge. The range $d \leq d_0$ where no instability is observed is shown in blue. The range $d_0 \leq d \leq d_1$ where a small perturbation of the front is observed without being followed by a significant growth of fingers is represented in green. Finally, the range $d \geq d_1$, in white, corresponds to the experiments where a fully developed fingering instability is observed. It happens that the diameter $d_f$ is systematically slightly smaller than the initial contact diameter $d_i$ ($d_f/d_i = 0.79 \pm 0.06$) and closely follows its evolution with the pillar diameter $d$ ($d_i = d$ for $d \leq d_2$ and $d_i = 600 \pm 100$ μm for $d \geq d_2$). From the values of $d_f$ and the count of the number $n_f$ of emerging fingers, we are able to compute the wavelength of the fingering instability at onset $\lambda_f = \pi d_f / n_f$ which appears to be nearly constant $\lambda_f = 104 \pm 17$ μm, independent of $d$ (see the inset in Fig. 10).

The fingering instability of a thin layer of elastic material confined between a plane and a spherical rigid substrates has been studied experimentally by Shull et al. in 2000.[25] The main results of this work is that the instability develops only when the contact diameter is larger than $5e$ and that the normalised wavelength $\lambda_f/e \simeq 4.5$ where $e$ is the elastic layer thickness. These two results are in good agreement with our experiments

Paper | Soft Matter
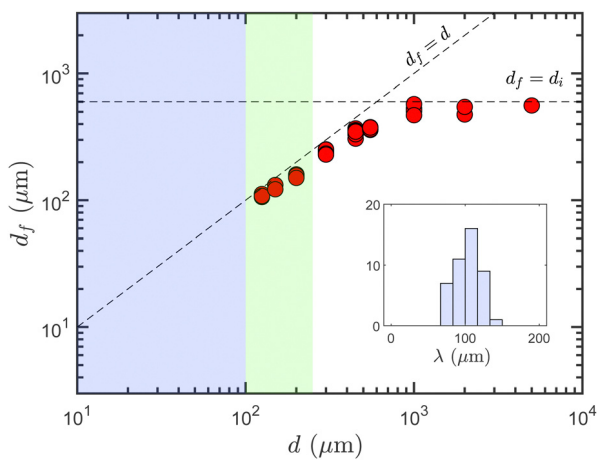

Fig. 10 Diameter of the contact $d_f$ at the onset of the fingering instability as a function of the pillar diameter $d$. Undulations of the debonding front are not observed for $d \leq d_0 \simeq 100$ μm (blue/left region). For $d_0 \leq d \leq d_1 = 225 \pm 25$ μm, a few undulations are observed. For $d \geq d_1$, a fully developed fingering instability is observed. The dashed lines recall the two asymptotic behaviours of the initial contact diameter $d_i$: $d_i = d$ for $d \leq d_2 = 775 \pm 225$ μm and $d_i = 600 \pm 100$ μm for $d \geq d_2$. The inset shows the histogram of the wavelengths $\lambda_f = \pi d_f/n_f$ calculated for all experiments whatever the pillar diameter.

for which $\lambda_f/e \simeq 4.7$ and the instability onset is observed only for contact diameters $d_f$ larger than 100 μm $\simeq 4.5 \times e$. These values of the wavelength of the fingering instability of a debonding front are also compatible with the one, $\lambda_f \simeq 95$ μm $\simeq 5e$, obtained by Lamblet et al.[4] during the peeling experiments of a PSA tape.

## 4 Discussion

### 4.1 The fingering instability

The fingering instability of the debonding (or healing) front of a confined thin layer of elastic material has been thoroughly studied theoretically[26,27] and experimentally.[25,28–30] In this section, we review this literature before we compare it to our experimental results involving a thin layer of a Pressure Sensitive Adhesive (PSA).

In ref. 28–30, where an elastic layer is confined between a rigid plate and a flexible plate submitted to a bending, Ghatak et al. show that the wavelength $\lambda_f$ of the perturbation of the debonding front is proportional to the adhesive layer thickness $e$, the ratio $\lambda_f/e$ being close to 4. The instability wavelength is more precisely proven to be purely geometric and to depend neither on the elastic modulus of the confined elastic film, on the bending stiffness of the cover plate, on the interfacial energy between the adhesive layer and the substrate, nor on the rate at which the debonding process is realised. Beside, the length of the fingers $L_f$ is shown to increase with the bending stiffness (via the associated increase in the radius of curvature) of the plate covering the confined elastic layer.

The results of Ghatak et al.[28,29] are in good agreement with the theories for the debonding of a confined incompressible elastic layer in the peeling geometry developed in ref. 26 and 27 which predict a wavelength of the debonding front proportional to the thickness of the elastic layer, with a geometric prefactor between 3 and 4. These theories, based on elastic fracture mechanics, reveal that it becomes energetically more favorable for the crack to grow with a crack front affected by a fingering instability when the elastic layer becomes thinner than a critical thickness proportional to the characteristic length $(\mu/D)^{1/3}$, where $D$ is the bending stiffness of the backing tape and $\mu$ the shear modulus of the elastic layer (for an incompressible material $\mu = E/3$ with $E$ the Young modulus). This prediction is in good agreement with the experimental results of Ghatak et al.[29] who reported a critical confinement parameter of $(D/\mu e^3)^{1/3} \simeq 18$ for the instability onset whereas Adda-Bedia and Mahadevan[26] predict theoretically a critical value of $(D/\mu e^3)^{1/3} \simeq 21$.

Vilmin et al.[27] go further in the theoretical analysis and predict that the length of the fingers should scale as $L_f \simeq (W_0 \rho^2 e/E)^{1/4}$ where $W_0$ is the work of adhesion per unit surface of debonded area and $\rho$ the typical radius of curvature of the portion of covering plate in contact with the stretched region of the adhesive layer. In the case where the cover plate is flexible, it happens that this expression can be rewritten in a simple way using the bending stiffness of the plate $D$ which scales as $D \propto W_0 \rho^2$ (when the crack front is progressing). It is also worth to note that the expression obtained here for the length of the fingers is identical to the one for the length over which the adhesive layer is stretched derived by Kaelble in 1960[8,15] in order to describe the peeling of an adhesive layer composed of individual elastic strands stretched between a rigid substrate and a flexible plate. In any case, this characteristic length results from the equilibrium between the curvature of the cover plate and the stretching of the adhesive elastic layer whose confinement has beforehand been lifted (by the fingering instability in the case of Vilmin et al.[27]). In parallel, various scaling laws for the length of the fingers $L_f$ have been empirically proposed based on experimental observations,[29–31] including a proportionality with the critical confinement thickness: $L_f \simeq 0.2 \times (D/\mu)^{1/3}$.[29] Nevertheless, the scaling law proposed theoretically by Vilmin et al.[27] has never been explicitly tested experimentally to the best of our knowledge and the question of the length of the fingers remains open.

In our experiments, the digitation wavelength at onset, $\lambda_f = 104 \pm 17$ μm, is equal to $(4.7 \pm 0.8) \times e$ where $e = 22$ μm is the initial thickness of the adhesive layer. Beside, we can estimate the length of the fingers predicted by the scaling law of Vilmin et al.,[27] $L_f \simeq (W_0 \rho^2 e/E)^{1/4} \simeq 155$ μm, where $W_0 \simeq 40$ mN m$^{-1}$ is the work of adhesion between the PDMS substrate and the adhesive tape estimated in Section 2.1, $E \simeq 2 \times 10^5$ Pa and $\rho = R = 1.15$ cm is the glass lens radius of curvature. The obtained value $L_f \simeq 155$ μm provides an order of magnitude compatible with the maximum finger lengths effectively observed in our experiments $L_f \simeq 120$ μm (see Fig. 9). Obviously, this consistency cannot be thought as a validation of the theoretical scaling law since none of the involved parameters has been varied experimentally, and further experimental works are needed. Nevertheless, despite the visco-elastic nature of the

 

adhesive layer we studied, our observations regarding the fingering instability are, at least to the first order, in good agreement with previous experimental and theoretical works on the digitation instability of a thin, confined, incompressible and purely elastic layer.

### 4.2 Relation between the critical elongation and the debonding force

In this section, we analyse the consistency between the power law experimentally observed for the critical elongation at debonding $\delta_c$ and the one observed for the maximum force during debonding $F_m$ for the regime at small pillar diameter, *i.e.* for $d \leq d_c \simeq 500$ μm.

In Section 3.2, we have seen that during the initial stage of the separation experiments at small pillar diameter, the stretched adhesive material follows an elastic behaviour. During this stage, the experimental evolution of the contact force $F$ with the elongation $\delta$ is well described by a simple Hooke's law

$$F = \frac{E_e \pi d^2}{4e} \delta, \qquad (1)$$

with $E_e \simeq 1.48 \times 10^5$ Pa an effective elastic modulus of the adhesive layer, independent of the pillar diameter $d$. Besides, we observe experimentally (see Fig. 9) that, still focusing on the experiments at small pillar diameter, the maximum force $F_m$ is typically reached when the elongation $\delta$ reaches half its critical value at debonding $\delta_c$. At this moment, the force–separation curve $F(\delta)$ has started to deviate from its initial linear elastic behavior (see Fig. 9). The maximum force $F_m$ is indeed typically 20% lower than the linear elastic extrapolation by eqn (1). This observation is actually still compatible with a linear elastic behavior if one considers that, at the moment $F = F_m$, the experimental contact area is typically 20% lower than the initial contact area $\pi d^2/4$. Combining these experimental facts, we can write the following approximate relation between the maximum force and the critical elongation

$$F_m \simeq 0.8 \frac{E_e \pi d^2}{8e} \delta_c. \qquad (2)$$

Relation (2) is perfectly consistent with the two power laws as a function of the pillar diameter $d$ evidenced for the maximum force $F_m$ and for the critical elongation at debonding $\delta_c$. Indeed, the experimentally observed power law $\delta_c \simeq (\ell_0 d)^{1/2}$ (Fig. 8) combined to eqn (2) leads to the relation

$$F_m \simeq 0.8 \frac{E_e \pi \ell_0^{1/2}}{8e} d^{5/2}, \qquad (3)$$

which nicely matches the scaling law $F_m \simeq F_\infty (d/d_c)^{5/2}$ observed in Fig. 6. Using the experimental values of $F_\infty \simeq 38.2$ mN, $d_c \simeq 500$ μm and $\ell_0 \simeq 10$ μm, one can moreover check the reasonable consistency of the two prefactors, $0.8 E_e \pi \ell_0^{1/2}/8e \simeq 6.7 \times 10^6$ N m$^{-1}$ and $F_\infty/d_c^{5/2} \simeq 6.8 \times 10^6$ N m$^{-1}$.

We would like to emphasize that the formulas put forward in this section simply aim at revealing the consistency between the two experimental scaling laws, $F_m \propto d^{5/2}$ and $\delta_c \propto d^{1/2}$, that we reported for the small pillar diameter regime. Actual theoretical modeling of our observations would involve understanding the physics of the critical elongation scaling law $\delta_c \simeq (\ell_0 d)^{1/2}$, and especially the physical meaning of the characteristic length $\ell_0$. This implies to consider the large-strain deformation processes and the fracture mechanics at the interface proceeding after the initial elastic stage during the debonding, both of which are well beyond the scope of the equations presented in this section.

## 5 Conclusion

In this article, we report an experimental study of the separation between an elastomeric cylindrical pillar of PDMS and a layer of an acrylic pressure sensitive adhesive deposited on a glass lens with a large radius of curvature.

For small pillar diameters, from 30 to 500 μm, the initial contact before separation covers the whole top surface of the pillar. The separation process then consists in the formation, the elongation and the final debonding of a fibril of adhesive material whose initial diameter is controlled by the pillar diameter $d$. We are therefore able to study the debonding process of a fibril of adhesive material, created from the adhesive layer, as a function of its initial diameter $d$. The main results of our experimental study is that the maximum force during debonding $F_m$, the critical force at debonding $F_c$ and the critical elongation at debonding of the adhesive fibril $\delta_c$ follow power laws with the initial fibril diameter, $F_{m,c} \propto d^{5/2}$ and $\delta_c \propto d^{1/2}$ respectively. The consistency between these two power laws follows from a quasi-elastic behaviour of the adhesive material observed until the maximum force is reached which typically occurs when the fibril elongation reaches about half the critical elongation at debonding.

For large pillar diameters ($d \geq 1$ mm), the initial contact between the PDMS pillar and the adhesive layer is smaller than the pillar diameter such that the situation resembles a sphere-plane contact sandwiching a thin layer of adhesive material. In this regime, the debonding scenario naturally becomes (almost) independent of the pillar diameter and the maximum force during debonding $F_m$, the critical force at debonding $F_c$ and the critical elongation at debonding $\delta_c$ are nearly constant.

Despite these two regimes are very distinct, a common phenomenon develops during the final stage of the debonding for all the experiments with initial contact diameter larger than typically 100 μm: the debonding front progresses through the growth of a fingering instability whose features match well the previous experimental observations and the theoretical predictions of the digitation instability in the case of a thin and confined layer of a purely elastic material.

The central result of our study is that the critical elongation at debonding $\delta_c$ of a micrometric fibril extracted from the studied adhesive layer follows the power law $\delta_c \simeq (\ell_0 d)^{1/2}$ with $d$ the fibril initial diameter and $\ell_0 \simeq 10$ μm a characteristic length. As discussed in the introduction, such debonding criterion of an adhesive fibril is a crucial element in the modelling of the energy release rate in peeling experiments. As shown by Chopin *et al.*,[16] once the large deformation stress–strain elongation rheology of the adhesive material is known, the missing element to have a predictive model for the peeling surface energy is indeed the elongation at debonding of the adhesive fibrils formed during peeling *via* cavitation or digitation processes. It is important to highlight here that, during peeling experiments of acrylic adhesives, the fibrils of adhesive material which are formed in the process zone ahead of the debonding front by cavitation[8,10] or digitation[4] have typical diameters of a few times the initial thickness of the adhesive layer which corresponds well to the range of adhesive fibril diameter over which we evidenced the power law for the critical elongation.

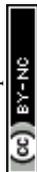

Nevertheless, before the adhesive fibril debonding criterion that we identified in our study, $\delta_c \simeq (\ell_0 d)^{1/2}$, could be applied to model the energy release rate in adhesive peeling experiments, two major questions have to be considered:

• In our experiments, the debonding front between the adhesive layer and the PDMS pillar is pinned on the circular edge of the pillar top surface during the first stage of the separation process. Is this feature a crucial ingredient in the observed scaling law for the critical elongation at debonding?

• What physical ingredients set the characteristic length $\ell_0$ driving the critical elongation of the adhesive material fibrils? These ingredients might include the interfacial work of adhesion between the adhesive and the substrate, the thickness of the adhesive layer, and the rate-dependant visco-elasticity and strain-hardening properties of the adhesive material, as it can be inferred from the peeling results of Villey *et al.*[8] and Chopin *et al.*[16]

These questions naturally call for new experimental works varying the interfacial energy, the rheology and thickness of the adhesive layer and the shape of the pillar that triggers the formation of the adhesive fibril.

## Conflicts of interest

There are no conflicts to declare.

## Acknowledgements

We thank L. Vanel, S. Santucci, L. Léger, and F. Restagno for helpful discussions. This work was supported by the Agence Nationale de la Recherche through Grant "AdhesiPS" No. ANR-17-CE08-0008.

## Notes and references